\documentclass[preprint]{aastex}

\received{2002 May 6}
\accepted{}

\slugcomment{The Astrophysical Journal, 2003}

\shorttitle{ORBITAL DECAY AND TIDAL DISRUPTION OF A STAR CLUSTER}
\shortauthors{MOURI \& TANIGUCHI}

\begin{document}

\title{ORBITAL DECAY AND TIDAL DISRUPTION OF A STAR CLUSTER: ANALYTICAL CALCULATION}

\author{HIDEAKI MOURI}
\affil{Meteorological Research Institute, Nagamine 1-1, Tsukuba 305-0052, Japan; hmouri@mri-jma.go.jp}

\and

\author{YOSHIAKI TANIGUCHI}
\affil{Astronomical Institute, Graduate School of Science, Tohoku University, Aoba, Sendai 980-8578, Japan; tani@astr.tohoku.ac.jp}

\begin{abstract}
The orbital decay and tidal disruption of a star cluster in a galaxy is studied in an analytical manner. Owing to dynamical friction, the star cluster spirals in toward the center of the galaxy. Simultaneously, the galactic tidal field strips stars from the outskirts of the star cluster. Under an assumption that the star cluster undergoes a self-similar evolution, we obtain the condition and timescale for the star cluster to reach the galaxy center before its disruption. The result is used to discuss the fate of so-called intermediate-mass black holes with $\ga 10^3$ $M_{\sun}$ found recently in young star clusters of starburst galaxies and also the mass function of globular clusters in galaxies.
\end{abstract}

\keywords{stellar dynamics --- 
          galaxies: star clusters ---
          galaxies: starburst}

\notetoeditor{You may find the expression ``$10^1$'' etc. in our order-of-magnitude discussion. Please do not replace it with, e.g., ``10''.}

\section{INTRODUCTION}

Tremaine, Ostriker, \& Spitzer (1975) studied a circular motion of a star cluster around the center of a galaxy (see also Binney \& Tremaine 1987, p. 427). The star cluster interacts with background stars of the galaxy via dynamical friction, loses its orbital kinetic energy, and thereby spirals in toward the galaxy center. Tremaine et al. (1975) approximated the mass density $\rho_{\rm b}$ of the background stars with that of the singular isothermal sphere,
\begin{equation}
\label{eq1}
\rho_{\rm b}(R) = \frac{v_{\rm b}^2}{4 \pi G R^2},
\end{equation}
where $R$ is the galactocentric radius, $v_{\rm b}$ is the circular velocity of the background stars, and $G$ is the gravitational constant. The cluster mass $M$ was set to be constant. The circular velocity of the star cluster was set to be $v_{\rm b}$. Then the rate of change of the galactocentric radius of the star cluster was obtained as
\begin{equation}
\label{eq2}
R\frac{dR}{dt} = -0.4276 \ln \Lambda_{\rm b} \frac{GM}{v_{\rm b}},
\end{equation}
where $\ln \Lambda_{\rm b}$ is the Coulomb logarithm. Solving equation (\ref{eq2}) with the initial condition $R = R_{\rm i}$ at $t = 0$, Tremaine et al. (1975) found that the star cluster reaches the galaxy center at the time $t = \tau_{\rm decay}$,
\begin{equation}
\label{eq3}
\tau_{\rm decay} = \frac{1.169}{\ln \Lambda_{\rm b}} 
                   \frac{R_{\rm i}^2 v_{\rm b}}{GM}.
\end{equation}
This timescale $\tau_{\rm decay}$ has been often used to study the evolution of a star cluster.

However, in practice, the star cluster is subject to the galactic tidal field. Stars belonging to the cluster have to lie within the tidal radius $r_{\rm t}$ from the cluster center (Spitzer 1987, p. 101; see also Capriotti \& Hawley 1996):
\begin{equation}
\label{eq4}
r_{\rm t} = \frac{2R}{3}
            \left[ \frac{M}{3M_{\rm b}(R)} \right] ^{1/3} 
            \left[ 1-\frac{R}{3M_{\rm b}(R)}\frac{dM_{\rm b}(R)}{dR} \right] ^{-1/3},
\end{equation}
where $M_{\rm b}(R)$ is the total mass of the background stars within the cluster's galactocentric radius $R$, i.e., $M_{\rm b}(R) = 4 \pi \int^R_0 \rho_{\rm b} R^2 dR$. If equation (\ref{eq1}) is used for the mass distribution $\rho_{\rm b}$, equation (\ref{eq4}) becomes
\begin{equation}
\label{eq5}
r_{\rm t} = \frac{2}{3}
            \left( \frac{G}{2v_{\rm b}^2}  \right) ^{1/3} M^{1/3} R^{2/3}.
\end{equation}
Thus, with a decrease of the galactocentric radius $R$, the tidal radius $r_{\rm t}$ becomes small. Stars are accordingly stripped from the outskirts of the cluster and become part of the background stars. Since the cluster mass becomes small, the dynamical friction becomes inefficient. In addition, if the star cluster is not sufficiently massive, it dissolves before it reaches the galaxy center.

This effect of tidal stripping was incorporated in some past studies of the orbital decay of a star cluster, but they were numerical studies with limited ranges of the parameters. A more general analytical study is desirable.

Here we conduct the analytical calculation for the first time. The star cluster is assumed to evolve in a self-similar manner (\S2). Then the rate of change of the cluster mass is obtained as a function of the galactocentric radius of the star cluster (\S3, eq. [\ref{eq14}]). Using this mass loss rate together with equation (\ref{eq2}) for the rate of change of the galactocentric radius of the star cluster, we derive the condition and timescale for the star cluster to reach the galaxy center before it dissolves (\S4, eqs. [\ref{eq27}] and [\ref{eq28}]). We discuss the validity of our assumptions and the astrophysical applications of our result (\S5). The latter discussion focusses on the fate of so-called intermediate-mass black holes with mass $\ga 10^3$ $M_{\sun}$ found in young star clusters of starburst galaxies and also on the mass function of globular clusters in galaxies.

\section{SELF-SIMILAR EVOLUTION}

The description of a star cluster requires at least two independent internal parameters, e.g., the cluster mass $M$ and the half-mass radius $r_{\rm h}$, in addition to external parameters such as the galactocentric radius $R$. To simplify our calculation, we consider a self-similar evolution, where one of the two internal parameters is exceptionally determined by the other. Here is an idealized self-similar model for a star cluster in a tidal field, which originates in the work of Spitzer (1987, p. 59; see also Capriotti \& Hawley 1996).

Recall that the crossing timescale $\tau_{\rm cross} \simeq (r_{\rm h}^3/GM)^{1/2}$ is much less than the other timescales relevant to the evolution of a star cluster. The star cluster is always nearly in virial equilibrium (Binney \& Tremaine 1987, p. 211):
\begin{equation}
\label{eq6}
E = K+W = -K = \frac{W}{2},
\end{equation}
where $E$ is the total energy, $K$ is the kinetic energy, and $W$ is the potential energy of the star cluster. Their definitions are
\begin{equation}
\label{eq7}
K = \frac{M \langle v^2 \rangle}{2}
\quad {\rm and} \quad
W = \frac{M \langle \Phi \rangle}{2},
\end{equation}
where $v$ is the velocity of a star, $\Phi$ is the gravitational potential at the position of a star, and $\langle \cdot \rangle$ denotes an average over stars in the cluster. Spitzer (1969) observed that various spherical clusters approximately have the relation
\begin{equation}
\label{eq8}
\langle v^2 \rangle = \frac{2GM}{5r_{\rm h}}.
\end{equation}
Then we have $E = -GM^2/5r_{\rm h}$, which yields
\begin{equation}
\label{eq9}
\frac{dE}{dt} = \frac{GM^2}{5r_{\rm h}^2} \frac{dr_{\rm h}}{dt}
               -\frac{2GM}{5r_{\rm h}} \frac{dM}{dt}.
\end{equation}
This change in total energy $dE$ is due to mass loss across the tidal radius $r_{\rm t}$ and is thereby equal to the change in potential energy associated with the displacement of $dM$ from the tidal radius to infinity. Thus we also have
\begin{equation}
\label{eq10}
\frac{dE}{dt} = - \frac{GM}{r_{\rm t}} \frac{dM}{dt}.
\end{equation}
Equations (\ref{eq9}) and (\ref{eq10}) combine to yield
\begin{equation}
\label{eq11}
\frac{d \ln r_{\rm h}}{dt}
=
3 \left( 2-5\frac{r_{\rm h}}{r_{\rm t}} \right) \frac{d \ln r_{\rm t}}{dt},
\end{equation}
where we have used $M \propto r_{\rm t}^3$ (eq. [\ref{eq5}]). If $r_{\rm h}/r_{\rm t} = 1/3$, this ratio remains the same. Since the tidal radius $r_{\rm t}$ is determined by the cluster mass $M$ and the galactocentric radius $R$ (eq. [\ref{eq5}]), the evolution of the star cluster is also determined by these two quantities.

\section{MASS LOSS RATE}

The escape of the individual stars from the cluster across the tidal radius is due to two-body relaxation. This process is thereby represented by removing a fraction $f_{\rm esc}$ of the stars every relaxation timescale $\tau_{\rm relax}$ (Binney \& Tremaine 1987, p. 490 and p. 523; Spitzer 1987, p. 52):
\begin{equation}
\label{eq12}
\frac{dM}{dt} = - \frac{f_{\rm esc}}{\tau_{\rm relax}} M.
\end{equation}
The half-mass relaxation timescale is used for $\tau_{\rm relax}$ (Spitzer \& Hart 1971; Binney \& Tremaine 1987, p. 514; Spitzer 1987, p. 40):
\begin{equation}
\label{eq13}
\tau_{\rm relax} = \frac{0.1386}{\ln \Lambda} 
                   \frac{M}{m}
                   \left( \frac{r_{\rm h}^3}{GM} \right)^{1/2},
\end{equation}
where $\ln \Lambda$ is the Coulomb logarithm for the star cluster and $m$ is the mean stellar mass. Using equation (\ref{eq5}), we rewrite equation (\ref{eq13}) as $\tau_{\rm relax} = (0.05335/\ln \Lambda) \times (r_{\rm h}/r_{\rm t})^{3/2} \times (MR/mv_{\rm b})$. Substituting this into equation (\ref{eq12}), we obtain
\begin{equation}
\label{eq14}
\frac{dM}{dt} 
= -18.74 
\ln \Lambda 
\left(\frac{r_{\rm h}}{r_{\rm t}} \right)^{-3/2} 
\frac{f_{\rm esc}mv_{\rm b}}{R}.
\end{equation}
If $r_{\rm h}/r_{\rm t}$ is constant, the mass loss rate in equation (\ref{eq14}) becomes larger with a decrease of the galactocentric distance $R$. This is because the tidal radius $r_{\rm t}$, the half-mass radius $r_{\rm h}$, and hence the relaxation timescale $\tau_{\rm relax}$ become smaller.

The escape probability $f_{\rm esc}$ is approximated with the fraction of stars in a Maxwellian distribution that have velocities exceeding the root-mean-square escape velocity $\langle v_{\rm esc}^2 \rangle^{1/2}$ (Binney \& Tremaine 1987, p. 490; Spitzer 1987, p. 51 and p. 57):
\begin{equation}
\label{eq15}
f_{\rm esc} = \frac{4 \pi}{(2 \pi)^{3/2}} 
              \int^{\infty}_{\sqrt{3\langle v_{\rm esc}^2 \rangle / \langle v^2 \rangle}}
              \exp \left( - \frac{v^2}{2} \right) v^2 dv.
\end{equation}
The escape velocity $v_{\rm esc}$ is defined at a position in the star cluster as
\begin{equation}
\label{eq16}
\frac{v_{\rm esc}^2(r)}{2} + \Phi (r) = - \frac{GM}{r_{\rm t}}.
\end{equation}
The average of equation (\ref{eq16}) is taken over positions of stars in the cluster. Using also equations (\ref{eq6})--(\ref{eq8}), we obtain 
\begin{equation}
\label{eq17}
\frac{\langle v_{\rm esc}^2 \rangle}{\langle v^2 \rangle} 
=
4- 5\frac{r_{\rm h}}{r_{\rm t}}.
\end{equation}
Thus the escape probability is constant in a self-similar model. The ratio $r_{\rm h}/r_{\rm t} = 1/3$ leads to $\langle v_{\rm esc}^2 \rangle / \langle v^2 \rangle = 7/3$, which in turn leads to $f_{\rm esc} = 0.07190$.

\section{ANALYTICAL SOLUTION}

With the initial condition $R = R_{\rm i}$ and $M = M_{\rm i}$ at $t = 0$, the behavior of a self-similar star cluster is studied by solving equations (\ref{eq2}) and (\ref{eq14}). The Coulomb logarithms $\ln \Lambda_{\rm b}$ and $\ln \Lambda$, the radius ratio $r_{\rm h}/r_{\rm t}$, and the escape probability $f_{\rm esc}$ are set to be constant. We introduce the friction timescale $\tau_{\rm fric}$ and the mass loss timescale $\tau_{\rm loss}$:
\begin{equation}
\label{eq18}
\tau_{\rm fric} = \left. \frac{R}{-dR/dt} \right| _{t=0}
                = \frac{R_{\rm i}^2 v_{\rm b}}{0.4276 \ln \Lambda_{\rm b} GM_{\rm i}},
\end{equation}
and
\begin{equation}
\label{eq19}
\tau_{\rm loss} = \left. \frac{M}{-dM/dt} \right| _{t=0}
                = \frac{(r_{\rm h}/r_{\rm t})^{3/2}M_{\rm i}R_{\rm i}}
                       {18.74 \ln \Lambda f_{\rm esc} m v_{\rm b}}.
\end{equation}
\notetoeditor{In equations (18) and (19), the right vertical bar indicates that the value is "at" t = 0.}
We also introduce the ratio between $\tau_{\rm fric}$ and $\tau_{\rm loss}$,
\begin{equation}
\label{eq20}
\alpha = \frac{\tau_{\rm fric}}{\tau_{\rm loss}}
       = 43.83 \frac{\ln \Lambda}{\ln \Lambda_{\rm b}} 
         \left( \frac{r_{\rm h}}{r_{\rm t}} \right)^{-3/2}
         \frac{f_{\rm esc}m v_{\rm b}^2 R_{\rm i}}{G M_{\rm i}^2}.
\end{equation}
The galactocentric radius of the cluster $R$, the cluster mass $M$, and the time $t$ are normalized as
\begin{equation}
\label{eq21}
\frac{R}{R_{\rm i}}       = \tilde{R}, \quad
\frac{M}{M_{\rm i}}       = \tilde{M}, \quad {\rm and} \quad
\frac{t}{\tau_{\rm fric}} = \tilde{t}.
\end{equation}
Equations (\ref{eq2}) and (\ref{eq14}) are accordingly rewritten as
\begin{equation}
\label{eq22}
\frac{d\tilde{R}}{d\tilde{t}} = - \frac{\tilde{M}}{\tilde{R}}
\quad {\rm and} \quad
\frac{d\tilde{M}}{d\tilde{t}} = - \frac{\alpha}{\tilde{R}}.
\end{equation}
The initial condition is $\tilde{R}_{\rm i} = \tilde{M}_{\rm i} = 1$ at $\tilde{t} = 0$.

Equation (\ref{eq22}) yields $\tilde{M}d\tilde{M}/d\tilde{t} = \alpha d\tilde{R}/d\tilde{t}$. The solution of this equation is
\begin{equation}
\label{eq23}
\tilde{M}^2 = 2\alpha \tilde{R} + 1-2\alpha.
\end{equation}
Using equation (\ref{eq23}), we eliminate $\tilde{M}$ from equation (\ref{eq22}):
\begin{equation}
\label{eq24}
\frac{d\tilde{t}}{d\tilde{R}} = - \frac{\tilde{R}}{(2\alpha\tilde{R}+1-2\alpha)^{1/2}}.
\end{equation}
The solution of equation (\ref{eq24}) gives the relation between the time $\tilde{t}$ and the galactocentric radius of the star cluster $\tilde{R}$:
\begin{equation}
\label{eq25}
\tilde{t} = \frac{(1-2\alpha-\alpha\tilde{R})
                  (1-2\alpha+2\alpha\tilde{R})^{1/2}-(1-3\alpha)}
                 {3\alpha^2}.
\end{equation}
The relation between the time $\tilde{t}$ and the cluster mass $\tilde{M}$ is obtained by substituting equation (\ref{eq23}) into equation (\ref{eq25}):
\begin{equation}
\label{eq26}
\tilde{t} = \frac{(3-6\alpha-\tilde{M}^2)\tilde{M}-2(1-3\alpha)}{6\alpha ^2}.
\end{equation}
Figure 1 illustrates the evolutions of $\tilde{R}$ and $\tilde{M}$ as a function of $\tilde{t}$ for $\alpha \rightarrow 0$ and $\alpha = 1/4$, 1/2, 3/4, and 1.

\placefigure{Fig1}

The condition for the star cluster to reach the galaxy center before its dissolution is $\tilde{M} \ge 0$ at $\tilde{R} = 0$, for which equation (\ref{eq23}) yields
\begin{equation}
\label{eq27}
\alpha \le \frac{1}{2} 
\quad {\rm or} \quad 
\frac{\tau_{\rm fric}}{\tau_{\rm loss}} \le \frac{1}{2}.
\end{equation}
The time at which the star cluster reaches the galaxy center is obtained by substituting $\tilde{R} = 0$ into equation (\ref{eq25}):
\begin{equation}
\label{eq28}
\tilde{\tau}_{\rm decay} 
=\frac{\tau_{\rm decay}}{\tau_{\rm fric}}
=\frac{(1-2\alpha)^{3/2}-(1-3\alpha)}{3\alpha^2}.
\end{equation}
If the parameter $\alpha$ is very small ($\alpha \ll 1/2$), equation (\ref{eq28}) is reduced to
\begin{equation}
\label{eq29}
\tilde{\tau}_{\rm decay} = \frac{1}{2} + \frac{\alpha}{6}
\quad {\rm or} \quad
\tau_{\rm decay} 
=
\frac{\tau_{\rm fric}}{2} \left( 1+\frac{\tau_{\rm fric}}{3\tau_{\rm loss}} \right).
\end{equation}
In the limit $\alpha \rightarrow 0$, equations (\ref{eq28}) and (\ref{eq29}) reproduce the solution $\tilde{\tau}_{\rm decay} = 1/2$ for no mass loss obtained by Tremaine et al. (1975, see our eq. [\ref{eq3}]). The timescale $\tilde{\tau}_{\rm decay}$ for $\alpha = 1/2$ is 2/3. Thus mass loss lengthens $\tilde{\tau}_{\rm decay}$ by a factor of 4/3 at most because significant mass loss occurs only at the latest stage of the orbital decay, i.e., in the vicinity of the galaxy center (Fig. 1). The cluster mass at the time when the star cluster reaches the galaxy center is obtained by substituting $\tilde{R} = 0$ into equation (\ref{eq23}):
\begin{equation}
\label{eq30}
\tilde{M}_{\rm decay}=(1-2\alpha)^{1/2}
\quad {\rm or} \quad
M_{\rm decay} = M_{\rm i} \left( 1-\frac{2\tau_{\rm fric}}{\tau_{\rm loss}} \right)^{1/2}.
\end{equation}
On the other hand, if the star cluster dissolves before it reaches the galaxy center, the corresponding time is obtained by substituting $\tilde{M} = 0$ into equation (\ref{eq26}):
\begin{equation}
\label{eq31}
\tilde{\tau}_{\rm dissolve} = \frac{1}{\alpha} -\frac{1}{3\alpha^2}
\quad {\rm or} \quad
\tau_{\rm dissolve} = \tau_{\rm loss} 
                      \left( 1-\frac{\tau_{\rm loss}}{3\tau_{\rm fric}} \right).
\end{equation}
The corresponding galactocentric radius is obtained by substituting $\tilde{M} = 0$ into equation (\ref{eq23}):
\begin{equation}
\label{eq32}
\tilde{R}_{\rm dissolve} = 1-\frac{1}{2\alpha}
\quad {\rm or} \quad
R_{\rm dissolve} = R_{\rm i} \left( 1-\frac{\tau_{\rm loss}}{2\tau_{\rm fric}} \right).
\end{equation} 
Therefore, the behavior of the star cluster is determined by the parameter $\alpha = \tau_{\rm fric}/\tau_{\rm loss}$. If $\alpha \ll 1/2$, the mass loss is slow (2$\tau_{\rm fric} \ll \tau_{\rm loss}$). The star cluster reaches the galaxy center before it loses much of its mass. If $\alpha \gg 1/2$, the mass loss is fast ($2\tau_{\rm fric} \gg \tau_{\rm loss}$). The star cluster dissolves before it moves significantly from its initial galactocentric radius.

\section{DISCUSSION}
\subsection{Model Assumptions}

We have assumed that a star cluster loses its mass via tidal stripping and two-body relaxation alone. This assumption is correct only if the orbit of the star cluster is circular and in the galactic plane. The star cluster otherwise suffers from gravitational shocks in passages through the galactic bulge or disk (Ostriker, Spitzer, \& Chevalier 1972). These complicated cases are beyond the reach of our analytical calculation.

We have assumed that the star cluster always has $r_{\rm h}/r_{\rm t} = 1/3$. This ratio is unstable in our idealized model (eq. [\ref{eq11}]). If $r_{\rm h}/r_{\rm t} > 1/3$, the ratio increases still more. The star cluster dissolves at $r_{\rm h}/r_{\rm t} = 4/5$ where the escape velocity $v_{\rm esc}$ is zero (Capriotti \& Hawley 1996; see our eq. [\ref{eq17}]). If $r_{\rm h}/r_{\rm t} < 1/3$, the ratio decreases toward zero. The ratio typical of globular and open clusters is $r_{\rm h}/r_{\rm t} \simeq 0.2$ (Binney \& Tremaine 1987, p. 26). Nevertheless, in practice, such star clusters do not undergo $r_{\rm h}/r_{\rm t} \rightarrow 0$. The ratio is rather sustained to be nearly constant by heating due to binaries, a process that is not considered in our idealized model (Binney \& Tremaine 1987, p. 543; Spitzer 1987, p. 148). Since the mass loss rate for $r_{\rm h}/r_{\rm t} = 0.2$ obtained from equations (\ref{eq14}), (\ref{eq15}), and (\ref{eq17}) is lower only by 12\% than that for $r_{\rm h}/r_{\rm t} = 1/3$, our model with $r_{\rm h}/r_{\rm t} = 1/3$ is of practical use.

There are other self-similar models for a star cluster in a tidal field. They are based on different assumptions. For example, H\'{e}non (1961) assumed the presence of an energy source at the cluster center and obtained $r_{\rm h}/r_{\rm t} = 0.1446$ and $f_{{\rm esc}} = 0.045$ (see also Spitzer 1987, p. 59).\footnote{
H\'{e}non's model was numerical. If we use our analytical approximation in equations (\ref{eq15}) and (\ref{eq17}), $r_{\rm h}/r_{\rm t} = 0.1446$ yields $f_{{\rm esc}} = 0.02006$.} 
The energy source could represent the binary heating and sustains the small value of $r_{\rm h}/r_{\rm t}$ as compared with that of our present model. It is possible to adapt these self-similar models into our calculation because the escape probability $f_{\rm esc}$ is constant in any self-similar model (eqs. [\ref{eq15}] and [\ref{eq17}]). We nevertheless favor our present model, which is simple but still sufficient for our purpose.

We have assumed that a star with velocity greater than the escape velocity immediately escapes from the cluster. With somewhat improved treatments of the escape condition, the escape probabilities $f_{\rm esc}$ close to ours were obtained by Fokker-Planck numerical simulations of Lee \& Ostriker (1987) and Lee \& Goodman (1995).

The escape probability $f_{{\rm esc}}$ is affected by two remarkable processes that are not considered in the above calculations. First, since stars with velocities slightly above the escape velocity can escape only through small holes near the Lagrangian points, they stay in the cluster for many crossing timescales. Some of the stars are scattered to lower velocities and become bound again. The escape probability is accordingly lowered (Baumgardt 2001). It might be better to consider that our $f_{{\rm esc}}$ value tends to be an upper limit. Second, when the star cluster is young, mass loss in the course of evolution of the individual stars, i.e., stellar wind and supernova explosion, causes shrink of the tidal radius and thereby enlarges the escape probability. This process is nevertheless unimportant in our study, which applies mainly to relatively old clusters.

Taking account of the distribution of stellar masses as well as the evolution of stars and binaries, Portegies Zwart et al. (2002) conducted realistic $N$-body numerical simulations of a star cluster in a static tidal field. While they did not vary the total number of stars in the cluster, they varied its initial density distribution and its distance from the galaxy center. The mass loss timescale $\tau_{\rm loss}$ was found to scale with the relaxation timescale at the tidal radius (eq. [\ref{eq13}] but with $r_{\rm t}$ substituted for $r_{\rm h}$).\footnote{
Within a static tidal field, the cluster mass decreases linearly with time (Spitzer 1987, p. 58). This behavior is reproduced by our mass loss rate (eq. [\ref{eq14}]), if the galactocentric radius of the star cluster is set to be constant.} 
Even if this is generally the case, our approach is still of use. We only have to set $f_{\rm esc} = \tau_{\rm relax}/\tau_{\rm loss}$ with $r_{\rm h}/r_{\rm t} = 1$.

\subsection{Astrophysical Applications}

The Coulomb logarithms $\ln \Lambda_{\rm b}$ and $\ln \Lambda$ are of order $10^1$, and the circular velocity $v_{\rm b}$ is of order $10^2$ km s$^{-1}$ (Binney \& Tremaine 1987, p. 423 and p. 427). Thus equations (\ref{eq18})--(\ref{eq20}) are written as
\begin{equation}
\label{eq33}
\tau_{\rm fric}
=
5.316 \times 10^9\ {\rm yr}\
\left( \frac{\ln \Lambda_{\rm b}}{10} \right)^{-1}
\left( \frac{v_{\rm b}}{100\ {\rm km}\ {\rm s}^{-1}} \right)
\left( \frac{M_{\rm i}}{10^6\ M_{\sun}} \right)^{-1}
\left( \frac{R_{\rm i}}{1\ {\rm kpc}} \right)^2,
\end{equation}
\begin{eqnarray}
\label{eq34}
\tau_{\rm loss}
 &=&
1.397 \times 10^{11}\ {\rm yr} 
\left( \frac{\ln \Lambda}{10} \right)^{-1}
\left( \frac{r_{\rm h}/r_{\rm t}}{1/3} \right)^{3/2}
\left( \frac{f_{\rm esc}}{0.07190} \right)^{-1}
\nonumber \\ & & \times
\left( \frac{m}{1\ M_{\sun}} \right)^{-1}
\left( \frac{v_{\rm b}}{100\ {\rm km}\ {\rm s}^{-1}} \right)^{-1}
\left( \frac{M_{\rm i}}{10^6\ M_{\sun}} \right)
\left( \frac{R_{\rm i}}{1\ {\rm kpc}} \right),
\end{eqnarray}
and
\begin{eqnarray}
\label{eq35}
\alpha 
&=& \frac{\tau_{\rm fric}}{\tau_{\rm loss}}
= 0.03807
\left( \frac{\ln \Lambda}{\ln \Lambda_{\rm b}} \right)
\left( \frac{r_{\rm h}/r_{\rm t}}{1/3} \right)^{-3/2}
\left( \frac{f_{\rm esc}}{0.07190} \right)
\nonumber \\ & & \times
\left( \frac{m}{1\ M_{\sun}} \right)
\left( \frac{v_{\rm b}}{100\ {\rm km}\ {\rm s}^{-1}} \right)^2
\left( \frac{M_{\rm i}}{10^6\ M_{\sun}} \right)^{-2}
\left( \frac{R_{\rm i}}{1\ {\rm kpc}} \right).
\end{eqnarray}
These formulae allow us to obtain the actual values of $\tau_{\rm decay}$, $M_{\rm decay}$, $\tau_{\rm dissolve}$, and $R_{\rm dissolve}$ (eqs. [\ref{eq28}], [\ref{eq30}], [\ref{eq31}], and [\ref{eq32}]). The condition for the star cluster to reach the galaxy center before its dissolution, $\alpha \le 1/2$ (eq. [\ref{eq27}]), is written as
\begin{eqnarray}
\label{eq36}
M_{\rm i} &\ge& 2.759 \times 10^5\ M_{\sun}\
\left( \frac{\ln \Lambda}{\ln \Lambda_{\rm b}} \right)^{1/2}
\left( \frac{r_{\rm h}/r_{\rm t}}{1/3} \right)^{-3/4}
\left( \frac{f_{\rm esc}}{0.07190} \right)^{1/2}
\nonumber \\ & & \times
\left( \frac{m}{1\ M_{\sun}} \right)^{1/2}
\left( \frac{v_{\rm b}}{100\ {\rm km}\ {\rm s}^{-1}} \right)
\left( \frac{R_{\rm i}}{1\ {\rm kpc}} \right)^{1/2}.
\end{eqnarray}
Figure 2 illustrates this condition as well as the values of $\tau_{\rm decay}$ and $\tau_{\rm dissolve}$ as a function of the initial mass $M_{\rm i}$ and the initial galactocentric radius $R_{\rm i}$.

\placefigure{Fig2}

\subsubsection{Fate of Intermediate-Mass Black Hole}

Our result is used to discuss the fate of massive black holes (BHs) found in young star clusters of starburst galaxies. At the 2 \micron\ secondary peak of the starburst galaxy M82, i.e., an active site of star formation, there is a source of compact X-ray emission (Matsushita et al. 2000; Kaaret et al. 2001; Matsumoto et al. 2001). The observed strong variability implies that the source is an accreting BH. The observed luminosity of $10^{41}$\,ergs\,s$^{-1}$ implies that the mass is greater than $10^3$\,$M_{\sun}$ if the emission is isotropic and its luminosity is below the Eddington limit. BHs of similar masses have been found also in other starburst galaxies, e.g., NGC 3628 (Strickland et al. 2001). They are called intermediate-mass BHs (Taniguchi et al. 2000) because their masses are in between those of stellar-mass BHs arising from stellar evolution and supermassive BHs found as the central engines of active galactic nuclei. The likely origin is successive mergers of massive stars and stellar-mass BHs that had segregated into the core of a star cluster (Taniguchi et al. 2000; Ebisuzaki et al. 2001; Portegies Zwart \& McMillan 2002). Under a favorable condition, timescales of these processes are less than $10^7$ yr (Mouri \& Taniguchi 2002b, c).

Owing to dynamical friction, the host cluster of an intermediate-mass BH sinks to the galaxy center. Upon dissolution, the cluster releases the BH. If there are more than one such BH, they merge with each other. Thus the intermediate-mass BH could evolve to be supermassive and serve as the central engine of an active galactic nucleus (Matsushita et al. 2000; Ebisuzaki et al. 2001).\footnote{
Two remarks. First, a single intermediate-mass BH is sufficient to explain the lowest-luminosity active galactic nuclei, where the mass of the central engine is as low as $10^2$ $M_{\sun}$ (e.g., NGC 4395; Filippenko \& Sargent 1989). Second, even if there remains no available gas when the intermediate-mass BH reaches the center, the galaxy has a chance to emerge as that with active galactic nucleus in the future, since supply of a large amount of gas to the nuclear region is repetitive (e.g., Mouri \& Taniguchi 2002a).} 
If this is the case, intermediate-mass BHs are crucial to studying a possible evolutionary connection between a starburst and an active galactic nucleus.

The host cluster is more massive than the BH and is affected by dynamical friction more significantly. Only if the host cluster reaches the galaxy center before it dissolves, the BH can reach the galaxy center in a reasonable timescale. If the host cluster dissolves before it moves significantly from its initial galactocentric radius, the timescale for the BH to reach the galaxy center is unreasonably large (Ebisuzaki et al. 2001).

Young star clusters in starburst galaxies have masses up to $10^6$ $M_{\sun}$ (e.g., Mengel et al. 2002). Largest-mass clusters contain largest numbers of massive stars and thus are most advantageous to form intermediate-mass BHs. Such clusters are also most advantageous to bring the BHs to the galaxy center. Star forming regions of starburst galaxies generally have galactocentric radii of $R \simeq 10^2$--$10^3$ pc (e.g., Telesco, Dressel, \& Wolstencroft 1993). If the mass is $10^6$ $M_{\sun}$, a star cluster there reaches the galaxy center within the timescale $10^8$--$10^9$ yr (Fig. 2). The timescale is $10^8$ yr for the host cluster of the intermediate-mass BH in M82, which lies 170 pc from the galaxy center. These timescales are very small as compared with the lifetime of a galaxy, $10^{10}$ yr.

\subsubsection{Mass Function of Globular Clusters}

Our result is also used to discuss the mass function of globular clusters in galaxies. The globular clusters have a preferred mass scale. Their mass function has a sharp peak at (1--$2) \times 10^5$ $M_{\sun}$ (e.g., Harris 1991). A plausible explanation is that the mass function was initially wide but later modified by dynamical friction, tidal stripping, gravitational shocks, and so on (e.g., Fall \& Rees 1977; Okazaki \& Tosa 1995; Fall \& Zhang 2001). Although the shape of the initial mass function is uncertain, it is possible to demonstrate that the cluster mass $10^5$ $M_{\sun}$ is advantageous for a star cluster to survive from dynamical friction and tidal stripping (see also Lee \& Ostriker 1987; Capriotti \& Hawley 1996).

Suppose that star clusters were formed $\tau_{\rm age}$ yr ago over ranges of the mass $M_{\rm i}$ and the galactocentric radius $R_{\rm i}$. On our $M_{\rm i}$--$R_{\rm i}$ diagram (Fig. 2), the star clusters that have survived up to now are those above the line for the timescale $\tau_{\rm age}$. This line is concave. If $R_{\rm i}$ is fixed, star clusters with too large masses have disappeared into the galaxy center due to dynamical friction. Star clusters with too small masses have dissolved due to tidal stripping. There is a initial mass $M_{\rm survive}$ with which star clusters have survived for the widest range of $R_{\rm i}$. This mass corresponds to $\alpha \simeq 1/2$ (solid line in Fig. 2), where the net effect of dynamical friction and tidal stripping is minimal. Eliminating $R_{\rm i}$ from equations (\ref{eq33}) and (\ref{eq35}) with $\tau_{\rm age} = \tau_{\rm decay} = \tau_{\rm dissolve} = 2\tau_{\rm fric}/3$ and $\alpha = 1/2$, we obtain
\begin{eqnarray}
\label{eq37}
M_{\rm survive} &=& 2.539 \times 10^5\ M_{\sun} 
                  \left( \frac{\ln \Lambda_{\rm b}}{10} \right) ^{-1/3}
                  \left( \frac{\ln \Lambda}{10} \right) ^{2/3}
                  \left( \frac{r_{\rm h}/r_{\rm t}}{1/3} \right)^{-1}
                  \left( \frac{f_{\rm esc}}{0.07190} \right) ^{2/3}
                  \nonumber \\ & & \times
                  \left( \frac{m}{1 M_{\sun}} \right) ^{2/3}
                  \left( \frac{v_{\rm b}}{100\ {\rm km}\ {\rm s}^{-1}} \right)
                  \left( \frac{\tau_{\rm age}}{10^{10}\ {\rm yr}} \right) ^{1/3}.
\end{eqnarray}
Thus, at the present age of globular clusters $\tau_{\rm age} \simeq 10^{10}$ yr, those with $M_{\rm i} \simeq 10^5$ $M_{\sun}$ are relatively abundant. Their present masses are also of order $10^5$ $M_{\sun}$ in most cases because mass loss is significant only at the latest stage of the orbital decay.

\acknowledgments
This work has been supported in part by the Japanese Ministry of Education, Science, and Culture under grants 10044052 and 10304013. The authors are grateful to the referee for helpful comments.

\section*{FIGURE CAPTIONS}

\figcaption[Fig1]{Evolutions of $\tilde{R} = R/R_{\rm i}$ ($a$) and $\tilde{M} = M/M_{\rm i}$ ($b$) for $\alpha = \tau_{\rm fric}/\tau_{\rm loss} \rightarrow 0$ and $\alpha = 1/4$, 1/2, 3/4, and 1 as a function of $\tilde{t} = t/\tau_{\rm fric}$. Here $R$ is the galactocentric radius of the star cluster, $R_{\rm i}$ is its initial value, $M$ is the cluster mass, $M_{\rm i}$ is its initial value, $t$ is the time, $\tau_{\rm fric}$ is the friction timescale (eq. [\ref{eq18}]), and $\tau_{\rm loss}$ is the mass loss timescale (eq. [\ref{eq19}]). The dotted lines show the $\tilde{R}$-$\tilde{t}$ relation for the dissolution of the star cluster ($a$) and the $\tilde{M}$-$\tilde{t}$ relation for the arrival of the star cluster at the galaxy center ($b$).}

\figcaption[Fig2]{Timescales $\tau_{\rm decay}$ with which the star cluster reaches the galaxy center ({\it dashed lines}, eq. [\ref{eq28}]) and $\tau_{\rm dissolve}$ with which the star cluster dissolves ({\it dotted lines}, eq. [\ref{eq31}]). The abscissa is the initial value of the cluster mass $M_{\rm i}$ in units of $M_{\sun}$. The ordinate is the initial value of the cluster's galactocentric radius $R_{\rm i}$ in units of pc. The other parameters are the same as in eqs. (\ref{eq33})--(\ref{eq36}). The solid line corresponds to $\alpha = \tau_{\rm fric}/\tau_{\rm loss} = 1/2$ (eqs. [\ref{eq35}] and [\ref{eq36}]) and divides the regions where the star cluster reaches the galaxy center before it dissolves ({\it right}) and the star cluster dissolves before it reaches the galaxy center ({\it left}).}


\begin{references}

\reference{} Baumgardt, H. 2001, \mnras, 325, 1323
\reference{} Binney, J., \& Tremaine, S. 1987, Galactic Dynamics (Princeton: Princeton Univ. Press)
\reference{} Capriotti, E. R., \& Hawley, S. L. 1996, \apj, 464, 765 (Erratum: \apj, 483, 984)
\reference{} Ebisuzaki, T., et al. 2001, \apjl, 562, L19
\reference{} Fall, S. M., \& Rees, M. J. 1977, \mnras, 181, 37p
\reference{} Fall, S. M., \& Zhang, Q. 2001, \apj, 561, 751
\reference{} Filippenko, A. V., \& Sargent, W. L. W. 1989, \apjl, 342, L11
\reference{} Harris, W. E. 1991, \araa, 29, 543
\reference{} H\'{e}non, M. 1961, Ann. d'Astrophys., 24, 369
\reference{} Kaaret, P., et al. 2001, \mnras, 321, L29
\reference{} Lee, H. M., \& Goodman, J. 1995, \apj, 443, 109
\reference{} Lee, H. M., \& Ostriker, J. P. 1987, \apj, 322, 123
\reference{} Matsumoto, H., et al. 2001, \apjl, 547, L25
\reference{} Matsushita, S., et al. 2000, \apjl, 545, L107
\reference{} Mengel, S., Lehnert, M. D., Thatte, N., \& Genzel, R. 2002, \aap, 383, 137
\reference{} Mouri, H., \& Taniguchi, Y. 2002a, \apj, 565, 786
\reference{} Mouri, H., \& Taniguchi, Y. 2002b, \apjl, 566, L17
\reference{} Mouri, H., \& Taniguchi, Y. 2002c, \apj, 580, in press (astro-ph/0208053)
\reference{} Okazaki, T., \& Tosa, M. 1995, \mnras, 274, 48
\reference{} Ostriker, J. P., Spitzer, L., Jr., \& Chevalier, R. A. 1972, \apjl, 176, L51 
\reference{} Portegies Zwart, S. F., Makino, J., McMillan, S. L. W., \& Hut, P. 2002, \apj, 565, 265
\reference{} Portegies Zwart, S. F., \& McMillan, S. L. W. 2002, \apj, 576, 899
\reference{} Spitzer, L., Jr. 1969, \apjl, 158, L139
\reference{} Spitzer, L., Jr. 1987, Dynamical Evolution of Globular Clusters (Princeton: Princeton Univ. Press)
\reference{} Spitzer, L., Jr., \& Hart, M. H. 1971, \apj, 164, 399
\reference{} Strickland, D. K., Colbert, E. J. M., Heckman, T. M., Weaver, K. A., Dahlem, M., \& Stevens, I. R. 2001, \apj, 560, 707
\reference{} Taniguchi, Y., Shioya, Y., Tsuru, T. G., \& Ikeuchi, S. 2000, \pasj, 52, 533
\reference{} Telesco, C. M., Dressel, L. L., \& Wolstencroft, R. D. 1993, \apj, 414, 120
\reference{} Tremaine, S. D., Ostriker, J. P., \& Spitzer, L., Jr. 1975, \apj, 196, 407


\end{references}
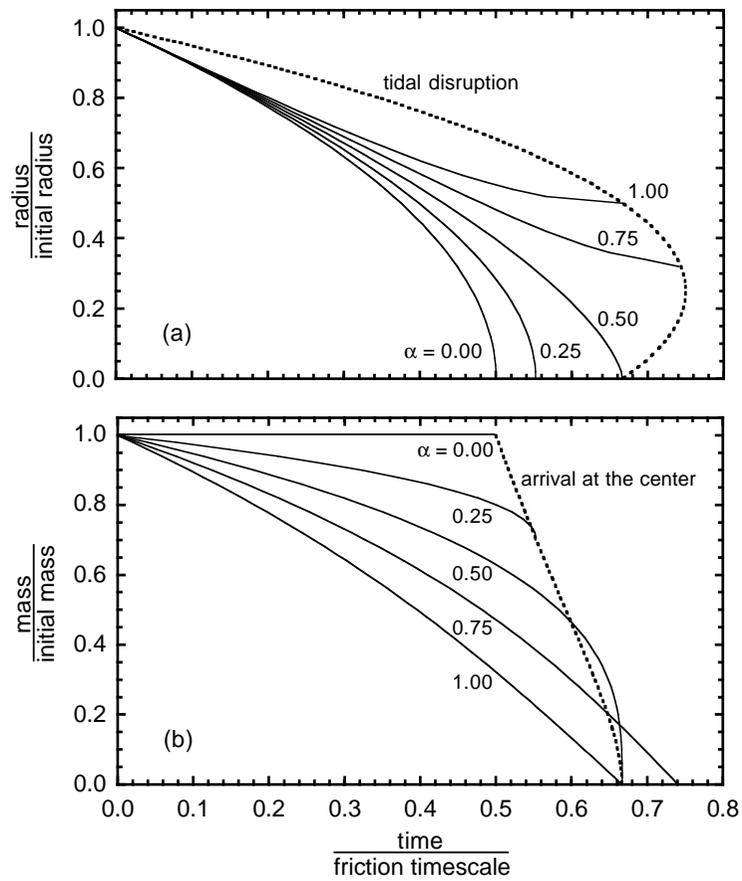
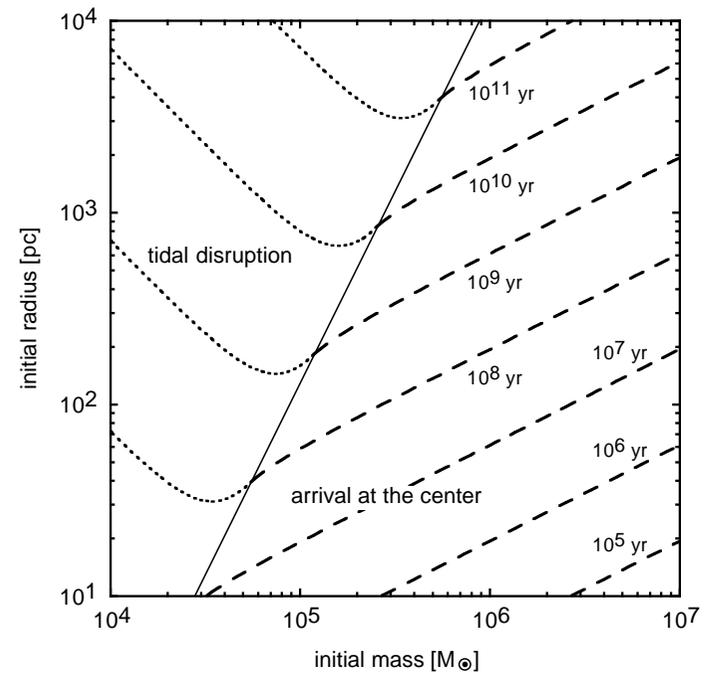
\end{document}